\documentstyle[12pt,epsfig]{article}     
\begin{document}


\def\beq             {\begin{equation}}
\def\eeq             {\end{equation}}
\def\beqd            {\begin{displaymath}}
\def\eeqd            {\end{displaymath}}
\def\baa             {\begin{array}}
\def\eaa             {\end{array}}
\def\beqaa           {\begin{eqnarray}}
\def\eeqaa           {\end{eqnarray}}
\def\beqaad          {\begin{eqnarray*}}
\def\eeqaad          {\end{eqnarray*}}
\def\btabu           {\begin{tabular}}
\def\etabu           {\end{tabular}}
\def\bfig            {\begin{figure}}
\def\efig            {\end{figure}}
\def\bce             {\begin{center}}
\def\ece             {\end{center}}

\def\ind             {\indent}
\def\noi             {\noindent}
\def\nn              {\nonumber}

\newcommand{\eq}[1]  {\mbox{eq.~(\ref{eq:#1})}}
\newcommand{\fig}[1] {\mbox{Fig.~\ref{fig:#1}}}   

\def\lbr             {\lbrack}
\def\rbr             {\rbrack}
\def\ti              {\tilde}
\def\q               {\bar}

\def\a               {\alpha}
\def\b               {\beta}
\def\d               {\delta}
\def\g               {\gamma}
\def\G               {\Gamma}
\def\l               {\lambda}
\def\t               {\theta}
\def\s               {\sigma}
\def\x               {\chi}

\def\sq              {\ti q}
\def\sqi             {\ti q_{i}}
\def\sqp             {\ti q^{\prime}}
\def\qp              {q^{\prime}}

\def\st              {\ti t}
\def\sti             {\ti t_{i}}

\def\sb              {\ti b}
\def\sbi             {\sb_{i}}

\def\ch              {\ti \x^\pm}
\def\chp             {\ti \x^+}
\def\chm             {\ti \x^-}
\def\nt              {\ti \x^0}

\newcommand{\msq}[1]   {m_{\sq_{#1}}}
\newcommand{\mst}[1]   {m_{\st_{#1}}}
\newcommand{\msb}[1]   {m_{\sb_{#1}}}
\newcommand{\mch}[1]   {m_{\ti \x^\pm_{#1}}}
\newcommand{\mnt}[1]   {m_{\ti \x^0_{#1}}}

\def\lij            {\ell_{ij}}
\def\kij            {k_{ij}}

\def\dl              {\d\ell}
\def\dk              {\dk}
\def\dth             {\d\theta}

\def\sg              {{\ti g}}
\def\hg              {^{(g)}}
\def\hsg             {^{(\sg )}}
\def\hsq             {^{(\sq )}}
\def\msg             {m_{\sg}}

\def\tW              {\t_W}
\def\tsq             {\t_{\sq}}
\def\sth             {\sin\t}
\def\cth             {\cos\t}
\def\sthq            {\sin^2\t}
\def\cthq            {\cos^2\t}

\def\Emiss           {E\llap/}

\def\onehf           {{\small\frac{1}{2} }} 
\def\oneth           {{\small\frac{1}{3} }}    
\def\twoth           {{\small\frac{2}{3} }}

\def\rzw             {\sqrt{2}}

\def\ra              {\rightarrow}

\def\BR              {\rm BR}
\def\gev             {\:{\rm GeV}}
\def\pb              {${\rm pb}^{-1}$}

\def\leerz           {\hspace{1cm}\\}

\newcommand{\gsim}{\;\raisebox{-0.9ex}
           {$\textstyle\stackrel{\textstyle >}{\sim}$}\;}
\newcommand{\lsim}{\;\raisebox{-0.9ex}{$\textstyle\stackrel{\textstyle<}
           {\sim}$}\;}

\begin{titlepage}

\begin{flushright}
UWThPh-1996-35 \\
HEPHY-PUB 646/96\\
hep-ph/9605412\\
\end{flushright}

\bce

\vspace{1.5cm}
\begin{LARGE} {\bf
  SUSY-QCD corrections \\[2mm]
  to scalar quark decays \\[2mm]
  into charginos and neutralinos \\[2mm]
}\end{LARGE}

\vspace{1.5cm}

\begin{large} 
  S. Kraml$^{\small 1}$, 
  H. Eberl$^{\small 1}$, \\[1mm]  
  A. Bartl$^{\small 2}$, 
  W. Majerotto$^{\small 1}$,  
  W. Porod$^{\small 2}$ \\[1mm]
\end{large}

\vspace{0.8cm}

{\em (1) Institut f\"ur Hochenergiephysik, 
         \"Osterreichische Akademie der Wissenschaften, \\
         Vienna, Austria}\\[1mm]
{\em (2) Institut f\"ur Theoretische Physik, Universit\"at Wien, 
         Vienna, Austria}

\vspace{1.5cm}

\begin{abstract}
We calculate the supersymmetric ${\cal O}(\alpha_s)$ QCD corrections to the
decays $\sqi\to\qp\,\ch_{j}$ ($i,j = 1, 2$) and $\sqi\to q\,\nt_{k}$ 
($k = 1,\ldots 4$) within the Minimal Supersymmetric Standard Model. 
In particular we consider the decays of squarks of the 
third generation, $\st_{i}$ and  $\sb_{i}$ ($i = 1,2$),  
where the left--right mixing must be taken into account.
The corrections turn out to be of about 10\%, except for higgsino--like 
charginos or neutralinos, where they can go up to 40\%.

\end{abstract}

\ece 
\end{titlepage}

\baselineskip=21pt   

\section {Introduction}

In supersymmetry (SUSY) \cite{haber} every quark has two scalar partners, 
the squarks $\sq_{L}$ and $\sq_{R}$. 
Quite generally, $\sq_{L}$ and $\sq_{R}$ mix giving the mass 
eigenstates $\sq_{1}$ and $\sq_{2}$ (with $m_{\sq_{1}} < m_{\sq_{2}}$).
Whereas the mixing can be neglected for the partners of the light
quarks, it can be important for the third generation due to the 
Yukawa coupling which is proportional to $m_{t}$ or $m_{b}$ 
\cite{ellis}. 

In the case that the gluino is heavier than the squarks,
the squarks have the decay modes 
  $\sq_{1,2} \to q'\,\ch_{j}$ ($j = 1,2$), and 
  $\sq_{1,2} \to q \,\nt_{k}$ ($k = 1, \ldots 4$),
where $\ch_{j}$ and $\nt_{k}$ denote the charginos and neutralinos, 
respectively.  
These decays were discussed so far only on the basis of 
tree--level calculations, for the partners of the light quarks 
in ref.~\cite{barnett} and with inclusion of the third generation 
($\st_{i}$, $\sb_{i}$, $i = 1,2$) in ref.~\cite{baer}. 
A detailed study of the $\st_{i}$ and $\sb_{i}$ decays can be found
in \cite{porod, sq3}. 

Here we calculate the ${\cal O}(\alpha_s)$ QCD corrections (including
the exchange of SUSY particles) to these decays within the 
Minimal Supersymmetric Standard Model (MSSM).
In particular, we do not neglect the masses of quarks and take into 
account $\sq_{L}$-$\sq_{R}$ mixing, so that the formulae are also
applicable to the decays $\st_{i} \to b\,\chp_{j}$, 
$\st_{i} \to t\,\nt_{j}$, and $\sb_{i} \to t\,\chm_{j}$, 
$\sb_{i} \to b\,\nt_{j}$. We work in the on--shell renormalization
scheme. For the  renormalization of the squark mixing angle we use the 
scheme introduced in \cite{helmut}, where we applied it to the 
case of $e^{+}e^{-} \to \sq_{i}\sq_{j}^{\ast}$. We also give a
numerical analysis of the QCD corrections to the decays 
$\st_{i}\to b\,\chm_{j}$ and $\st_{i}\to t\,\nt_{k}$.  

For the decays of a squark into a light quark ($m_{q} = 0$) 
and a photino 
the SUSY-QCD corrections were already calculated in \cite{hikasa}.
The QCD corrections to the decay $t\to \st_{i}\,\nt_{k}$ 
have been computed very recently within the MSSM in ref.~\cite{djouadi}. 
The SUSY-QCD corrections 
to the strong decays $\sq\to q\,\sg$ were calculated in 
ref.~\cite{zerwas}.

\section {Tree level formulae}

The squarks $\sq_{L}$ and $\sq_{R}$ are related to the mass eigenstates 
$\sq_{1}$ and $\sq_{2}$ by:
\beq
  {\sq_1 \choose \sq_2} = {\cal R}^{\sq}\,{\sq_L \choose \sq_R},
  \hspace{4mm}
  {\cal R}^{\sq} = \left(\baa{ll} \cth_{\sq} & \sth_{\sq} \\ 
                                 -\sth_{\sq} & \cth_{\sq} \eaa\right) .
\eeq

\noi Their interaction with charginos $\ch_{j}$ ($j = 1,2$) and 
neutralinos $\nt_{k}$ ($k = 1, \ldots 4$)  is given by \cite{porod}:
\beq
  {\cal L} = g\,\bar{q}\, 
    (a_{ik}^{\sq}\,P_{R} + b_{ik}^{\sq}\,P_{L})\,\nt_{k}\,\sq_{i} + 
  g\,\bar{t}\, 
    (\ell_{ij}^{\sb}\,P_{R} + k_{ij}^{\sb}\,P_{L})\,\chp_{j}\,\sb_{i} +
  g\,\bar{b}\, 
    (\ell_{ij}^{\st}\,P_{R} + k_{ij}^{\st}\,P_{L})\,\x^{+ c}_{j}\,\st_{i} +
  {\rm h.c.}
  \label{eq:Lint}
\eeq
$q$, $t$, $b$, $\chp_{j}$, and $\nt_{k}$ denote the four--component 
spinors of the corresponding particles.
\noi The respective decay widths at tree level are 
\beq
  \Gamma^{0} (\sq_{i}\to q^{\prime}\ch_{j}) =
  \frac{g^{2}\lambda^{\onehf}(\msq{i}^{2}, m_{q^{\prime}}^{2}, \mch{j}^{2})}
       {16\pi\msq{i}^{3}}\,
  \Big( [(\ell_{ij}^{\sq})^{2} + (k_{ij}^{\sq})^{2}]\,X 
         - 4\,\lij^{\sq}k_{ij}^{\sq} m_{q^{\prime}}\mch{j} 
  \Big)
  \label{eq:chwidth}
\eeq
and
\beq
  \Gamma^{0} (\sq_{i}\to q\nt_{k}) =
  \frac{g^{2}\lambda^{\onehf}(\msq{i}^{2}, m_{q}^{2}, \mnt{k}^{2})}
       {16\pi\msq{i}^{3}} \,
  \Big( [(a_{ik}^{\sq})^{2} + (b_{ik}^{\sq})^{2}]\,X'
         - 4\,a_{ik}^{\sq} b_{ik}^{\sq} m_{q}\mnt{k} 
  \Big)
  \label{eq:ntwidth}
\eeq
where $\lambda (x,y,z) = x^{2}+y^{2}+z^{2}-2xy-2xz-2yz$ and 
\beq
  X = \msq{i}^{2} - m_{q^{\prime}}^{2} - \mch{j}^{2}, \hspace{4mm}
  X' = \msq{i}^{2} - m_{q}^{2} - \mnt{k}^{2}.
\label{eq:X}
\eeq
The $\sqi$-$q'$-$\ch_{j}$ couplings $\lij^{\sq}$ and $\kij^{\sq}$ are 
\beq
  \lij^{\sq} = {\cal R}^{\sq}_{in}\, {\cal O}^{q}_{jn},
  \hspace{4mm}
  \kij^{\sq} = {\cal R}^{\sq}_{i1}\, {\cal O}^{\qp}_{j2}
\eeq
with
\beq
  {\cal O}^{t}_{j} = { -V_{j1} \choose Y_{t}\,V_{j2} } ,
  \hspace{6mm}
  {\cal O}^{b}_{j} = { -U_{j1} \choose Y_{b}\,U_{j2} } .
\eeq

\noi
The $\sqi$-$q$-$\nt_{k}$ couplings $a_{ik}^{\sq}$ and $b_{ik}^{\sq}$ 
are given by
\beq
  a_{ik}^{\sq} = {\cal R}^{\sq}_{in}\, {\cal A}^{q}_{kn},
  \hspace{4mm}
  b_{ik}^{\sq} = {\cal R}^{\sq}_{in}\, {\cal B}^{q}_{kn}
\eeq
with
\beq
  {\cal A}^{q}_{k} = { f_{Lk}^{q} \choose h_{Rk}^{q} },
  \hspace{6mm}
  {\cal B}^{q}_{k} = { h_{Lk}^{q} \choose f_{Rk}^{q} } ,
\eeq
and
\begin{small}
\beq \begin{array}{ll}
  h_{Lk}^{t} = ~Y_{t} \left( 
               N_{k3}\sin\b - N_{k4}\cos\b \right),  &
  f_{Lk}^{t} = -\frac{2\sqrt{2}}{3} \sin\tW N_{k1} - \sqrt{2}\,
        (\onehf - \twoth\sin^{2}\tW ) \frac{N_{k2}}{\cos\tW}, \\
  h_{Rk}^{t} = ~Y_{t} \left( 
               N_{k3}\sin\b - N_{k4}\cos\b \right),  &
  f_{Rk}^{t} = -\frac{2\sqrt{2}}{3} \sin\tW 
               (\tan\tW N_{k2} - N_{k1}), 
  \label{eq:hfstop}
\end{array} \eeq
\beq \begin{array}{ll}
  h_{Lk}^{b} = -Y_{b} \left( 
               N_{k3}\cos\b + N_{k4}\sin\b \right), &
  f_{Lk}^{b} = \frac{\sqrt{2}}{3}\sin\tW N_{k1} +
     \sqrt{2}\,(\onehf - \oneth\sin^{2}\tW ) \frac{N_{k2}}{\cos\tW}, \\
  h_{Rk}^{b} = -Y_{b} \left( 
               N_{k3}\cos\b + N_{k4}\sin\b   \right),  &
  f_{Rk}^{b} = \frac{\sqrt{2}}{3}\sin\tW 
                   (\tan\tW N_{k2} - N_{k1}).                            
  \label{eq:hfsbot}
\end{array} \eeq
\end{small}

\noi 
$N_{ij}$ is the $4\times 4$ unitary matrix diagonalizing the 
neutral gaugino--higgsino mass matrix in the basis $\ti\gamma$, 
$\ti Z^{0}$, $\ti H^{0}_{1}\cos\b - \ti H^{0}_{2}\sin\b$,
$\ti H^{0}_{1}\sin\b + \ti H^{0}_{2}\cos\b$ \cite{haber, neutralinos}. 
$U_{ij}$ and $V_{ij}$ are the $2\times 2$ unitary matrices diagonalizing 
the charged gaugino--higgsino mass matrix \cite{haber, charginos}. 
Assuming CP conservation, we choose a phase convention 
in which $N_{ij}$, $U_{ij}$, and $V_{ij}$ are real. 
$Y_{f}$ denotes the Yukawa coupling,
\beq
  Y_{t} = m_{t}/(\sqrt{2}\,m_{W}\sin\b), \hspace{6mm} 
  Y_{b} = m_{b}/(\sqrt{2}\,m_{W}\cos\b).
  \label{eq:yukcop}
\eeq

\section {SUSY--QCD corrections}

The ${\cal O}(\a_{s})$ SUSY-QCD corrected decay width 
can be decomposed in the following way:
\beq
  \G = \G^{0} + \d\G^{(v)} + \d\G^{(w)} + \d\G^{(c)} + \d\G_{g,real}.
  \label{eq:gencorr}
\eeq
The superscript $v$ denotes the vertex correction (Figs. 1b, c), 
$w$ the wave function correction (Figs. 1d-h), 
and $c$ the shift from the bare to the on-shell couplings. 
$\d\G_{g,real}$ is the correction due to real gluon emission 
(Figs. 1i, 1j) which has to be included in order to achieve infrared 
finiteness. 
According to \eq{chwidth} $\d\G^{(a)}$ ($a = v,\,w,\,c$) can be written as
\beqaa
\d\G^{(a)} (\sq_{i}\to q^{\prime}\ch_{j}) &=&
  \frac{g^{2}\lambda^{\onehf}(\msq{i}^{2}, m_{q^{\prime}}^{2}, \mch{j}^{2})}
       {16\pi\msq{i}^{3}} \\
 & &\hspace*{3mm}
  \left[ (2\,\lij^{\sq}\,\d\lij^{\sq\,(a)} 
        + 2\,\kij^{\sq}\,\d\kij^{\sq\,(a)})\,X          
    - 4\,m_{q'}\mch{j} (\lij^{\sq}\,\d\kij^{\sq\,(a)} 
                      + \kij^{\sq}\,\d\lij^{\sq\,(a)}) 
  \right] \nn
\eeqaa
with $X$ as defined in \eq{X}.
An analogous expression holds for $\d\G^{(a)} (\sq_{i}\to q\,\nt_{k})$ 
by replacing $\ch_{j} \to \nt_{k}$, $q' \to q$, $X\to X'$, 
$\ell_{ij}^{\sq} \to a_{ik}^{\sq}$, $k_{ij}^{\sq} \to b_{ik}^{\sq}$, 
$\d\ell_{ij}^{\sq\,(a)} \to \d a_{ik}^{\sq\,(a)}$, and 
$\d k_{ij}^{\sq\,(a)} \to \d b_{ik}^{\sq\,(a)}$.
$\d\lij^{\sq\,(a)}$, $\d k_{ij}^{\sq\,(a)}$,  etc. 
get contributions from gluon exchange, 
gluino exchange, and the four--squark interaction. 
As we will see, in the renormalization scheme used  
the four--squark interaction contribution turns out to be zero. \\
In what follows we will take the squark decays into charginos as example. 
For the decays $\sq_{i}\to q\,\nt_{k}$ one has to make the 
replacements just mentioned before.

\subsection{Vertex corrections}

The gluonic vertex correction (Fig. 1b) yields
\beqaa
\d\ell_{ij}^{\sq\,(v,g)} &=& \frac{\alpha_{s}}{3\pi} 
  \Big\{
  \big[ (4m_{q'}^{2} + 2X) (C_{0}+C_{1}+C_{2}) +
         (2\mch{j}^{2} + X) C_{1} + B_{0} \big]\,\ell_{ij}^{\sq} \nn \\
  & & \hspace{13mm} + \big[ 2m_{q'}\mch{j}C_{2} \big]\,k_{ij}^{\sq}
  \Big\}, \\
\d k_{ij}^{\sq\,(v,g)} &=& \frac{\alpha_{s}}{3\pi} 
  \Big\{
  \big[ (4m_{q'}^{2} + 2X) (C_{0}+C_{1}+C_{2}) +
         (2\mch{j}^{2} + X) C_{1} + B_{0} \big]\,k_{ij}^{\sq} \nn \\
  & & \hspace{13mm} + \big[ 2m_{q'}\mch{j}C_{2} \big]\,\ell_{ij}^{\sq}
  \Big\}.   
\eeqaa
$B_{0}$, $C_{0}$, $C_{1}$, and $C_{2}$ are the standard 
two-- and three--point functions \cite{pave}. 
In this case, $B_{0} = B_{0}(\mch{j}^{2}, \msq{i}^{2}, m_{q'}^{2})$ 
and $C_{m} = C_{m}(\msq{i}^{2}, \mch{j}^{2}, m_{q'}^{2};
\l^{2}, \msq{i}^{2}, m_{q'}^{2})$,
where we follow the conventions of \cite{denner}.
As usually, we introduce a gluon mass $\l$ for the regularization of 
infrared divergencies. 
 
The contribution to the vertex correction due to the graph shown in 
Fig. 1c with a gluino and a squark $\ti q'_{n}$ $(n = 1, 2)$ 
in the loop is:
\beqaa
\d{\ell_{ij}^{\sq}}^{(v,\sg)} &=& \frac{2}{3}\frac{\alpha_{s}}{\pi} 
  \Big\{
  \mch{j}\big[ (m_{q'}\a_{LR}+m_{q}\a_{RL}-\msg\a_{LL})\,\ell_{nj}^{\sq'} 
            + \mch{j}\a_{RL} k_{nj}^{\sq'} \big] C_{1} \nn \\
  & &+ m_{q'}\big[ (m_{q'}\a_{RL}-m_{q}\a_{LR}+\msg\a_{RR}) k_{nj}^{\sq'} 
            - \mch{j}\a_{LR}\,\ell_{nj}^{\sq'} \big] (C_{1}+C_{2})  \\ 
  & &+ \msg\big[ (m_{q'}\a_{RR}-m_{q}\a_{LL}+\msg\a_{RL}) k_{nj}^{\sq'} 
            - \mch{j}\a_{LL}\,\ell_{nj}^{\sq'} \big] C_{0}
     + (X C_{1} + B_{0})\,\a_{RL} k_{nj}^{\sq'} 
  \Big\}, \nn  \\[2mm]     
\d{k_{ij}^{\sq}}^{(v,\sg)} &=& \frac{2}{3}\frac{\alpha_{s}}{\pi} 
  \Big\{
  \mch{j}\big[ (m_{q'}\a_{RL}+m_{q}\a_{LR}-\msg\a_{RR}) k_{nj}^{\sq'} 
            + \mch{j}\a_{LR}\,\ell_{nj}^{\sq'} \big] C_{1} \nn \\
& &+ m_{q'}\big[ (m_{q'}\a_{LR}-m_{q}\a_{RL}+\msg\a_{LL})\,\ell_{nj}^{\sq'} 
            - \mch{j}\a_{RL} k_{nj}^{\sq'} \big] (C_{1}+C_{2})  \\                        
& &+ \msg\big[ (m_{q'}\a_{LL}-m_{q}\a_{RR}+\msg\a_{LR})\,\ell_{nj}^{\sq'} 
            - \mch{j}\a_{RR} k_{nj}^{\sq'} \big] C_{0} 
     + (X C_{1} + B_{0})\,\a_{LR}\,\ell_{nj}^{\sq'}
  \Big\} \nn      
\eeqaa
with 
\beq \begin{array}{ll}
  \a_{LL} = \left( \a_{LL}\right)_{in} = 
    {\cal R}^{\sq}_{i1}\,{\cal R}^{\sq '}_{n1}, & 
  \a_{LR} = \left( \a_{LR}\right)_{in} = 
    {\cal R}^{\sq}_{i1}\,{\cal R}^{\sq '}_{n2}, \\   
  \a_{RL} = \left( \a_{RL}\right)_{in} = 
    {\cal R}^{\sq}_{i2}\,{\cal R}^{\sq '}_{n1}, &  
  \a_{RR} = \left( \a_{RR}\right)_{in} = 
    {\cal R}^{\sq}_{i2}\,{\cal R}^{\sq '}_{n2}.   
\end{array} \eeq

\noi Here, 
$B_{0} = B_{0}(\mch{j}^{2}, m_{\sq'_{n}}^{2}, m_{q}^{2})$, and
$C_{m} = C_{m}(\msq{i}^{2}, \mch{j}^{2}, m_{q'}^{2};
\msg^{2}, m_{q}^{2}, m_{\sq'_{n}}^{2})$. \\

\subsection{Wave--function correction}

The wave--function correction is given by
\beqaa
  \d\lij^{\sq\,(w)} &=& {\tiny\frac{1}{2}}\, \big[ 
    \d Z^{L^{\dagger}}_{q'} + \d\ti Z_{ii} \big]\,\lij^{\sq} 
    + \d\ti Z_{ii'}\,\ell_{i'j}^{\sq}, \\
  \d\kij^{\sq\,(w)} &=& {\tiny\frac{1}{2}}\, \big[ 
    \d Z^{R^{\dagger}}_{q'} + \d\ti Z_{ii} \big]\,\kij^{\sq} 
    + \d\ti Z_{ii'}\,k_{i'j}^{\sq}.                       
\eeqaa
$Z_{q'}^{L,R}$ are the quark wave--function renormalization constants 
due to gluon exchange (Fig. 1d),
\beq
  \d Z^{L^{\dagger}(g)}_{q'} = \d Z^{R^{\dagger}(g)}_{q'} =
  - \frac{2}{3}\frac{\alpha_{s}}{\pi} 
    \left[\, B_{0} + B_{1} 
           - 2m_{q'}^{2} (\dot B_{0} - \dot B_{1}) - r/2 \,\right] 
\label{eq:qwg}
\eeq            
with $B_{m} = B_{m}(m_{q'}^{2}, \l^{2}, m_{q'}^{2})$,
$\dot B_{m} = \dot B_{m}(m_{q'}^{2}, \l^{2}, m_{q'}^{2})$, 
and due to gluino exchange (Fig. 1e), 
\beq 
  \d Z^{L^{\dagger}(\sg)}_{q'} = \frac{2}{3}\frac{\alpha_{s}}{\pi}\,
  \Big\{ \cos^{2}\t_{\ti q'}B_{1}^{1} + \sin^{2}\t_{\ti q'}B_{1}^{2}  
         + m_{q'}^{2}\big[ \dot B_{1}^{1} + \dot B_{1}^{2} 
            + \frac{\msg}{m_{q'}} \sin 2\t_{\ti q'}
              (\dot B_{0}^{1} - \dot B_{0}^{2}) \big] \Big\},  
\eeq
\beq           
  \d Z^{R^{\dagger}(\sg)}_{q'} = \frac{2}{3}\frac{\alpha_{s}}{\pi}\,
  \Big\{ \sin^{2}\t_{\ti q'}B_{1}^{1} + \cos^{2}\t_{\ti q'}B_{1}^{2} 
         + m_{q'}^{2}\big[ \dot B_{1}^{1} + \dot B_{1}^{2} 
            + \frac{\msg}{m_{q'}} \sin 2\t_{\ti q'}
              (\dot B_{0}^{1} - \dot B_{0}^{2}) \big] \Big\},  
\eeq  
where $B_{m}^{i} = B_{m}(m_{q'}^{2},\,\msg^{2},\,m_{\ti q'_{i}}^{2})$ and 
$\dot B_{m}^{i} = \dot B_{m}(m_{q'}^{2},\,\msg^{2},\,m_{\ti q'_{i}}^{2})$.  
The parameter $r$ in \eq{qwg} and \eq{dmqg} exhibits the dependence 
on the regularization. As $r$ does not cancel in the f\/inal 
result we have to use the dimensional reduction scheme \cite{siegel} 
($r = 0$) which preserves supersymmetry at least at one--loop order. 

\noi
The squark wave--function renormalization constants $\ti Z_{in}$ 
stem from gluon, gluino, and squark loops according to Figs.~1f -- 1h . 
They are given by:
\beq
  \delta \ti Z_{ii}^{(g,\sg)} = 
    - \mbox{Re}\left\{\dot\Sigma_{ii}^{(g,\sg)}(\msq{i}^{2}) \right\}\,, 
  \quad
  \delta \ti Z_{ii'}^{(\sg,\sq)} =
  -\,\frac{\mbox{Re}\left\{\Sigma_{ii'}^{(\sg,\sq)}(\msq{i}^{2})\right\} 
         }{\msq{i}^2 - \msq{i'}^2} \,, 
  \quad i \neq i' 
\eeq
with $\dot\Sigma_{ii} (m^2) = 
\partial\Sigma_{ii} (p^2)/\partial p^2 |_{p^2=m^2}$. 
The squark self--energy contribution
due to gluon exchange (Fig. 1f) is
\beqaa  
  \dot\Sigma_{ii}^{(g)} (\msq{i}^2) &=& - \frac{\alpha_s}{3\pi} 
  \Big\{ 
    3B_{0}(\msq{i}^{2}, 0, \msq{i}^{2}) 
    + 2B_{1}(\msq{i}^{2}, 0, \msq{i}^{2}) \nn \\
  & & \hspace*{1cm}
    + 4\msq{i}^{2} \dot B_{0} (\msq{i}^{2}, \l^{2}, \msq{i}^{2})
       + 2\msq{i}^{2} \dot B_{1} (\msq{i}^{2}, 0, \msq{i}^{2}) 
  \Big\} \nn ,
\eeqaa
and that due to gluino exchange (Fig. 1g) is
\beqaa  
  \dot\Sigma_{ii}^{(\sg)} (\msq{i}^2) &=& \frac{2}{3}\frac{\a_{s}}{\pi} 
  \left[ B_0(\msq{i}^2, \msg^2,m_q^2) + (\msq{i}^2-m_q^2-\msg^2) 
         \dot B_0 (\msq{i}^2, \msg^2, m_q^2) \right.\nn\\
  & & \left.\hspace*{1cm} 
    -2 m_q\msg (-1)^i\sin 2\tsq \dot B_0 (\msq{i}^2, \msg^2,m_q^2) 
  \right] ,
\eeqaa
\beq
  \Sigma_{12}^{(\sg)}(\msq{i}^2) = \Sigma_{21}^{(\tilde g)}(\msq{i}^2) =
  \frac{4}{3}\frac{\alpha_{s}}{\pi} \msg m_q \cos 2 \tsq 
  B_0(\msq{i}^2, \msg^2,m_q^2).
\eeq
The four--squark interaction (Fig. 1h) gives
\beq
\Sigma_{12}\hsq (\msq{i}^2) = \Sigma_{21}\hsq (\msq{i}^2) = 
  \frac{\alpha_s}{6\pi}\sin 4\tsq
  \left[ A_{0}(m_{{\sq}_2}^2) - A_{0}(m_{{\sq}_1}^2) \right]
\eeq
where $A_{0}(p^{2})$ is the standard one-point function 
in the convention of \cite{denner}. 
Note that $\Sigma_{ii'}\hsq (p^2)$ is independent of $p^{2}$ and hence 
$\Sigma_{ii'}\hsq (\msq{1}^2) = 
 \Sigma_{ii'}\hsq (\msq{2}^2) = 
 \Sigma_{ii'}\hsq$.

\subsection{Renormalization of the bare couplings}

In order to make the shift from the bare to the on-shell couplings 
it is necessary to renormalize the quark mass as well as 
the squark mixing angle:
\beq 
  \d\lij^{\sq\,(c)} = 
    {\cal S}_{in}^{\sq}\,{\cal O}_{jn}^{q}\,\d\tsq + 
    {\cal R}^{\sq}_{i2}\,\d{\cal O}^{q}_{j2},
  \hspace{4mm}
  \d\kij^{\sq\,(c)} = 
    {\cal S}_{i1}^{\sq}\,{\cal O}_{j2}^{q'}\,\d\tsq +
    {\cal R}^{\sq}_{i1}\,\d{\cal O}^{q'}_{j2},
\eeq
\beq 
  \d a_{ik}^{\sq\,(c)} = 
    {\cal S}_{in}^{\sq}\,{\cal A}_{kn}^{q}\,\d\tsq +
    {\cal R}^{\sq}_{i2}\,\d h_{Rk}^{q}, 
  \hspace{4mm}
  \d b_{ik}^{\sq\,(c)} = 
    {\cal S}_{in}^{\sq}\,{\cal B}_{kn}^{q}\,\d\tsq + 
    {\cal R}^{\sq}_{i1}\,\d h_{Lk}^{q},
\eeq
where
\beq
  {\cal S}^{\sq}\d\tsq = \d\,{\cal R}^{\sq} =
  \left(\baa{ll} -\sth_{\sq} &  \cth_{\sq} \\ 
                 -\cth_{\sq} & -\sth_{\sq} \eaa\right) \d\tsq .
\eeq

\noi 
$\d{\cal O}^{t}_{j2} = 
V_{j2}/(\sqrt{2}\,m_{W}\sin\b)\,\d m_{t}$,
$\d{\cal O}^{b}_{j2} = 
U_{j2}/(\sqrt{2}\,m_{W}\cos\b)\,\d m_{b}$,
and analogously for $\d h_{Lk}^{q}$ and $\d h_{Rk}^{q}$ 
according to eqs.~(10) -- (12). 
The gluon contribution to $\d m_{q}$ is
\beq
  \d m_{q}\hg = - \frac{2}{3}\frac{\alpha_{s}}{\pi}\, 
    m_{q} \left[ B_{0} - B_{1} - r/2 \right] ,
  \quad B_{m} = B_{m}(m_{q}^{2}, 0, m_{q}^{2}),
\label{eq:dmqg}
\eeq
and the gluino contribution is
\beq
  \d m_{q}^{(\ti g)} = - \frac{\alpha_{s}}{3\pi} \left[ 
    m_{q} (B_{1}^{1} + B_{1}^{2})
    + \msg \sin 2\tsq (B_{0}^{1}-B_{0}^{2}) \right] ,
  \quad
  B_{m}^{i} = B_{m}(m_{q}^{2}, \msg , \msq{i}^{2}).
\eeq

\noi
For the renormalization of the squark mixing angle we use 
the scheme introduced in \cite{helmut}:
\beq  
  \delta\tsq\hsq = \frac{\alpha_s}{6\pi} \,
  \frac{\sin 4\tsq}{\msq{1}^2 - \msq{2}^2}
  \left[ A_{0}(\msq{2}^2) - A_{0}(\msq{1}^2) \right] .
\eeq
\beq  
  \delta\tsq\hsg = \frac{\alpha_s}{3\pi}
  \frac{m_{\sg} m_q}{I^{3L}_q (m_{\sq_1}^2 - m_{\sq_2}^2)}
  \left[ B_{0}(m_{\sq_2}^2,m_{\sg}^2,m_q^2)\,\ti v_{11} -
         B_{0}(m_{\sq_1}^2,m_{\sg}^2,m_q^2)\,\ti v_{22}\right] .
\eeq

\noi
with $\ti v_{ih}$ the $Z\sq_{i}\sq_{h}^{*}$ couplings,
$\ti v_{11}= 4(I_q^{3L} \cos^2 \tsq - s_W^2 e_q)$ and 
$\ti v_{22}= 4(I_q^{3L} \sin^2 \tsq - s_W^2 e_q)$. 
Here $I_q^{3L}$ is the third component of the weak isospin 
and $e_{q}$ the charge of the quark $q$. 

\noi With this choice of $\d\tsq$ the squark contribution
to the correction is zero: $\d\G^{(w,\sq)} + \d\G^{(c,\sq)} = 0$.
Moreover, the off-diagonal contribution $(i \neq h)$ of Fig.~1g 
vanishes in this scheme. 
We checked analytically that the resulting SUSY-QCD corrected 
decay width is ultraviolet f\/inite. 

\subsection{Real gluon emission}

The ${\cal O}(\a_{s})$ contribution from 
real gluon emission, as shown in Figs.~1i and 1j, gives 
the decay width of $\sq_{i}\to g\,q'\,\ch_{j}$:
\beqaa
  \G(\sq_{i}\to g q' \ch_{j}) &=& 
  -\frac{g^{2}\a_{s}}{6\pi^{2}\msq{i}}\, \Big\{
  \big[ (\kij^{\sq})^{2} + (\lij^{\sq})^{2} \big] (I_{1}^{0}+I) + \nn\\
  & & \hspace*{8mm}
  2 Z \big[ \msq{i}^{2} I_{00} + m_{q'}^{2} I_{11} + 
  (\msq{i}^{2}+m_{q'}^{2}-\mch{j}^{2}) I_{01} + I_{0} + I_{1} \big] 
  \Big\}
\eeqaa
where $Z = \big[ (\kij^{\sq})^{2} + (\lij^{\sq})^{2} \big] X - 
4\,\kij^{\sq} \lij^{\sq} m_{q'} \mch{j}$. 
The phase space integrals $I$, $I_{n}$, $I_{nm}$, and $I_{n}^{m}$ 
have $(\msq{i}, m_{q'}, \mch{j})$ as arguments and are 
given in \cite{denner}.
An analogous expression holds for the $\sq_{i}\to g\,q\,\nt_{k}$
decay width.

\section {Numerical results and discussion}

We now turn to the numerical analysis of the ${\cal O}(\a_{s})$ 
SUSY-QCD corrected decay widths. As examples we consider the decays 
$\st_{1} \to b\,\chp_{1}$ and $\st_{1} \to  t\,\nt_{1}$. 
Masses and couplings of the charginos and neutralinos depend on 
the parameters $M$, $M'$, $\mu$, and $\tan\b$.
The squark sector is determined by the soft--breaking parameters 
$M_{Q}$, $M_{U}$, and $M_{D}$, the trilinear couplings 
$A_{t}$ and $A_{b}$, and $\mu$ and $\tan\b$, which all enter the 
squark mass matrices. For the following analysis we 
use physical squark masses and mixing angles as input 
parameters.
We take $m_{t} = 180$ GeV, $m_{b} = 5$ GeV, 
$m_{Z} = 91.187$ GeV, $\sin^{2}\t_{W} = 0.23$, 
$\a_{w}(m_{Z}) = 1/128.87$, and
$\a_{s}(m_{Z}) = 0.12$. 
Moreover, we use the GUT relations 
$M' = \frac{5}{3}\,M\,\tan^{2}\t_{W} \sim 0.5 M$ and
$\msg = \frac{\a_{s}}{\a_{2}}\,M \sim 3 M$. 
Furthermore, we use $\a_{s}(Q^{2}) = 
4\pi/(b_{0}\,x)\,[ 1 - 2\,b_{1} \ln x /(b_{0}^{2}\,x)]$ with
$b_{0} = 11 - 2/3\,n_{f}$, $b_{1} = 51 - 19/3\,n_{f}$, and 
$x = \ln (Q^{2}/\Lambda^{2})$ \cite{PDG}. 
Here, $n_{f}$ is the number of f\/lavours. 

For discussing the $\st_{1}\to b\,\chp_{1}$ decay mode we take
$\mch{1} = 75$ GeV, $\mst{2} = 300$ GeV, $\msb{1} = 220$ GeV, 
$\msb{2} = 230$ GeV, $\cth_{\sb} = -0.4$, and $\tan\b = 2$.
In order to study the dependence on the nature of the chargino 
(gaugino-- or higgsino--like), 
we choose three sets of $M$ and $\mu$ values: 
$M \ll |\mu|$ ($M =  66$ GeV,  $\mu = -500$ GeV) , 
$M \sim |\mu|$ ($M =  70$ GeV,  $\mu =  -61$ GeV), and 
$M \gg |\mu|$ ($M = 300$ GeV,  $\mu =  -62$ GeV). 
In Fig.~2 we show the dependence of the QCD corrections on the 
$\st_{1}$ mass (in \% of the tree--level decay width) in the range 
of $\mst{1} = 80$ GeV (LEP2, Tevatron) to $\mst{1} = 220$ GeV 
(LHC, $e^{+}e^{-}$ linear collider). 
Here, we take $\cth_{\st} = 0.72$ which corresponds to $M_{Q}\sim M_{U}$.
If the $\chp_{1}$ is gaugino--like (solid line) the correction 
is about 10\% to 20\% for a light $\st_{1}$ ($\mst{1} \lsim 100$ GeV) 
and decreases with the stop mass. In the case of $M \sim |\mu|$ 
(dashed line) the correction lies between $\sim +10\%$ and $-10\%$.
Due to the large top Yukawa coupling 
the biggest ef\/fect is found for a higgsino--like $\chp_{1}$ 
(dash-dotted line). Here, the correction lies in 
the range of -10\% to -20\%. 

\noi
The $\cth_{\st}$ dependence is shown in Fig.~3 where we plot 
the ${\cal O}(\a_{s})$ corrected decay widths (solid lines) together
with the tree--level widths (dashed lines) as a function of 
$\cth_{\st}$ for $\mst{1} = 150$ GeV and the other parameters as 
given above. The widths show the characteristic dependence on the stop 
mixing angle: If $\st_{1} \sim \st_{R}$ ($\cth_{\st} \sim 0$) 
it strongly couples to the higgsino component of $\chp_{1}$, 
whereas for $\st_{1} \sim \st_{L}$ ($\cth_{\st} \sim \pm 1$) it 
strongly couples to the gaugino component. 
Again, a striking effect can be seen for $|\mu| \ll M$. 
However, also for $M \ll |\mu|$ and $M \sim |\mu|$ the correction
can be larger then 10\%, especially if the tree--level decay width 
is very small.

\noi
We have also studied the dependence on the masses and the mixing 
angle of the exchanged sbottom (Fig.~1c). 
In the cases studied these effects turn out to be small.
The dependence on the gluino mass is more complex: On the one hand, 
the gluino mass enters in the propagator of the graphs in 
Figs.~1c, 1e, and 1g. 
On the other hand, the mass and the couplings of $\chp_{1}$ in the 
final state also depend on $m_{\sg}$ via the relation $M \sim 0.3\,m_{\sg}$.
However, if one relaxes this relation keeping $M$ fixed and 
varying $m_{\sg}$ the correction increases with the gluino mass. 
This has also been noticed in refs.~\cite{hikasa, djouadi}.

For the decay of $\st_{1}$ into $t\,\nt_{1}$ we take 
$\mnt{1} = 50$ GeV, $\mst{2} = 400$ GeV, $\msb{1} = 320$ GeV, 
$\msb{2} = 340$ GeV, $\cth_{\sb} = -0.4$, and $\tan\b = 2$.
Again, we choose scenarios with a gaugino--like $\nt_{1}$
($M =  98$ GeV, $\mu = -500$ GeV) and a higgsino--like $\nt_{1}$ 
($M = 250$ GeV, $\mu =  -55$ GeV), 
as well as  one where $M \sim |\mu|$ ($M =  93$ GeV, $\mu =  -90$ GeV).  
We show in Fig.~4 the dependence of the SUSY-QCD correction 
(in \% of the tree--level width) on the $\st_{1}$ mass 
in the range of $\mst{1} = 230$ to 400 GeV, for $\cth_{\st} = 0.72$. 
As in the case of the $\st_{1}\to b\,\chp_{1}$ decay, the decay 
width gets the biggest correction if $\nt_{1}$ is higgsino--like: 
It is between -15\% and -20\% and decreases with increasing $\st_{1}$ mass. 
For $M \ll |\mu|$ ($M \sim |\mu|$) the correction lies between 
$\sim +8\%$ and $\sim -5\%$ ($\sim -3\%$).  
 
\noi The dependence of the tree--level (dashed lines) and the QCD 
corrected (solid lines) decay widths of $\st_{1}\to t\,\nt_{1}$ 
on the stop mixing angle is shown in Fig.~5  for 
$\mst{1} = 260$ GeV and the other parameters as given in Fig.~4.
Again there is a strong dependence on the nature of the lightest
neutralino. The corrections are large for a higgsino--like neutralino.


Summarizing, we have computed the ${\cal O}(\a_{s})$ SUSY-QCD 
corrections to the decays  of squarks into charginos and neutralinos.
We have concentrated on the third generation squarks, where the 
quark masses cannot be neglected, and the left--right mixing plays an 
essential r\^ole.
The corrections strongly depend on the nature of the charginos or 
neutralinos. They are about 10\%, but go up to 40\% for 
higgsino--like charginos/neutralinos.

\section*{Acknowledgements}

This work arose from the Workshop on Physics at LEP2, CERN 1995, and 
the Workshop on Physics with $e^{+}e^{-}$ Linear Colliders, 
Annecy -- Gran Sasso -- Hamburg, 1995. 
This work was supported by the ``Fonds zur F\"orderung der 
wissenschaftlichen Forschung'' of Austria, project no. P10843--PHY.
We thank A.~Djouadi for valuable correspondence which helped us find a
mistake in the computer program.

\newpage

\newpage
\section*{Figure captions}

\newcounter{fig}
\begin{list}{\bf Fig. \arabic{fig}:}{\usecounter{fig} 
  \labelwidth1.6cm \leftmargin1.8cm \labelsep0.4cm 
  \parsep0.8ex plus0.2ex minus0.1ex \itemsep0ex plus0.2ex}

\item Feynman diagrams relevant for the ${\cal O}(\a_{s})$ SUSY-QCD 
  corrections to squark decays into charginos and neutralinos:
  (a) tree level, (b) gluon vertex correction, (c) gluino vertex 
  correction, (d) and (e) quark wave--function renormalization, (f) and 
  (g) squark wave--function renormalization, (h) four--squark 
  interaction, and (i) and (j) real gluon emission.  \\
  
\item SUSY-QCD corrections for the decay $\st_{1}\to b\,\chp_{1}$ 
  in percent of the tree-level width as a function of $\mst{1}$ 
  for $\mch{1} = 75$ GeV, $\mst{2} = 300$ GeV, 
  $\msb{1} = 220$ GeV, $\msb{2} = 230$ GeV, $\cth_{\st} = 0.72$, 
  $\cth_{\sb} = -0.4$, $\tan\b = 2$. 
  Three scenarios are studied: 
  $M =  66$ GeV,  $\mu = -500$ GeV (solid line), 
  $M =  70$ GeV,  $\mu =  -61$ GeV (dashed line), and 
  $M = 300$ GeV,  $\mu =  -62$ GeV (dash-dotted line). \\

\item Tree--level (dashed lines) and SUSY-QCD corrected (solid lines) 
  widths of the decay $\st_{1}\to b\,\chp_{1}$ in GeV as a function 
  of $\cth_{\st}$ 
  for $\mch{1} = 75$ GeV, $\mst{1} = 150$ GeV, $\mst{2} = 300$ GeV, 
  $\msb{1} = 220$ GeV, $\msb{2} = 230$ GeV, $\cth_{\sb} = -0.4$, 
  $\tan\b = 2$. Three scenarios are studied:  
  (a) $M =  66$ GeV,  $\mu = -500$ GeV, 
  (b) $M =  70$ GeV,  $\mu =  -61$ GeV, and 
  (c) $M = 300$ GeV,  $\mu =  -62$ GeV. \\

\item SUSY-QCD corrections for the decay $\st_{1}\to t\,\nt_{1}$ 
  in percent of the tree-level width as a function of $\mst{1}$ 
  for $\mnt{1} = 50$ GeV, $\mst{2} = 500$ GeV, 
  $\msb{1} = 320$ GeV, $\msb{2} = 340$ GeV, $\cth_{\st} = 0.72$, 
  $\cth_{\sb} = -0.4$, $\tan\b = 2$.  
  Three scenarios are studied: 
  $M =  98$ GeV, $\mu = -500$ GeV (solid line), 
  $M =  93$ GeV, $\mu =  -90$ GeV (dashed line), and 
  $M = 250$ GeV, $\mu =  -55$ GeV (dash-dotted line).  \\

\item Tree--level (dashed lines) and SUSY-QCD corrected (solid lines) 
  widths of the decay $\st_{1}\to t\,\nt_{1}$ in GeV as a function 
  of $\cth_{\st}$ 
  for $\mnt{1} = 51$ GeV, $\mst{1} = 260$ GeV, $\mst{2} = 500$ GeV, 
  $\msb{1} = 320$ GeV, $\msb{2} = 340$ GeV, $\cth_{\sb} = -0.4$, 
  $\tan\b = 2$. Three scenarios are studied: 
  (a) $M =  98$ GeV, $\mu = -500$ GeV,
  (b) $M =  93$ GeV, $\mu =  -90$ GeV, and
  (c) $M = 250$ GeV, $\mu =  -55$ GeV.  \\

\end{list}


\newpage
\setlength{\unitlength}{1mm}
\begin{center}
\begin{picture}(135,210)
\put(0,15){\mbox{\psfig{file=fig1.eps,height=195mm}}}
\put(60,175){\makebox(0,0)[t]{\large{\bf{Fig.~1a}}}}
\put(39,193.5){\makebox(0,0)[r]{$\sq_i$}}
\put(83,211){\makebox(0,0)[l]{$q'\,(q)$}}
\put(83,177){\makebox(0,0)[l]{$\ch_{j}\,(\nt_{k})$}}

\put(25,125){\makebox(0,0)[t]{\large{\bf{Fig.~1b}}}}
\put(7,142.5){\makebox(0,0)[r]{$\sq_i$}}
\put(50,159.5){\makebox(0,0)[l]{$q'\,(q)$}}
\put(50,126.5){\makebox(0,0)[l]{$\ch_{j}\,(\nt_{k})$}}
\put(35,150){\makebox(0,0)[br]{$g$}}
\put(35,137){\makebox(0,0)[tr]{$\sq_i$}}
\put(42,142.5){\makebox(0,0)[l]{$q'\,(q)$}}
\put(92,125){\makebox(0,0)[t]{\large{\bf{Fig.~1c}}}}
\put(72,142.5){\makebox(0,0)[r]{$\sq_i$}}
\put(117.5,159.5){\makebox(0,0)[l]{$q'\,(q)$}}
\put(117.5,126.5){\makebox(0,0)[l]{$\ch_{j}\,(\nt_{k})$}}
\put(102,150){\makebox(0,0)[br]{$\sg$}}
\put(102,137){\makebox(0,0)[tr]{$q$}}
\put(109,142.5){\makebox(0,0)[l]{$\sq_n'\,(\sq_n)$}}
\put(13,90){\makebox(0,0)[t]{\large{\bf{Fig.~1d}}}}
\put(-2,97){\makebox(0,0)[r]{$q$}}
\put(29.5,97){\makebox(0,0)[l]{$q$}}
\put(13,106){\makebox(0,0)[b]{$g$}}
\put(13,95.5){\makebox(0,0)[t]{$q$}}
\put(13,63.5){\makebox(0,0)[t]{\large{\bf{Fig.~1e}}}}
\put(-2,71.5){\makebox(0,0)[r]{$q$}}
\put(29.5,71.5){\makebox(0,0)[l]{$q$}}
\put(13,79){\makebox(0,0)[b]{$\sg$}}
\put(13,69){\makebox(0,0)[t]{$\sq_n$}}
\put(61.5,90){\makebox(0,0)[t]{\large{\bf{Fig.~1f}}}}
\put(45,97){\makebox(0,0)[r]{$\sq_i$}}
\put(77,97){\makebox(0,0)[l]{$\sq_i$}}
\put(62,106){\makebox(0,0)[b]{$g$}}
\put(62,95.5){\makebox(0,0)[t]{$\sq_i$}}
\put(61.5,63.5){\makebox(0,0)[t]{\large{\bf{Fig.~1g}}}}
\put(45,71.5){\makebox(0,0)[r]{$\sq_i$}}
\put(77,71.5){\makebox(0,0)[l]{$\sq_h$}}
\put(62,79){\makebox(0,0)[b]{$\sg$}}
\put(62,69){\makebox(0,0)[t]{$q$}}
\put(110,79){\makebox(0,0)[t]{\large{\bf{Fig.~1h}}}}
\put(94.5,84){\makebox(0,0)[r]{$\sq_i$}}
\put(125,84){\makebox(0,0)[l]{$\sq_h$}}
\put(109.5,97){\makebox(0,0)[b]{$\sq_{1,2}$}}
\put(25,13){\makebox(0,0)[t]{\large{\bf{Fig.~1i}}}}
\put(7,31.5){\makebox(0,0)[r]{$\sq_i$}}
\put(50,49){\makebox(0,0)[l]{$q'\,(q)$}}
\put(50,15){\makebox(0,0)[l]{$\ch_{j}\,(\nt_{k})$}}
\put(30,41){\makebox(0,0)[l]{$g$}}
\put(92,13){\makebox(0,0)[t]{\large{\bf{Fig.~1j}}}}
\put(72,31.5){\makebox(0,0)[r]{$\sq_i$}}
\put(117.5,49){\makebox(0,0)[l]{$q'\,(q)$}}
\put(117.5,15){\makebox(0,0)[l]{$\ch_{j}\,(\nt_{k})$}}
\put(117,40){\makebox(0,0)[l]{$g$}}
\end{picture}

\setcounter{figure}{1}


\vspace{1cm}
\noi
\begin{minipage}[b]{125mm} {\setlength{\unitlength}{1mm} 
\begin{picture}(130,85)
\put(6,4){\mbox{\epsfig{figure=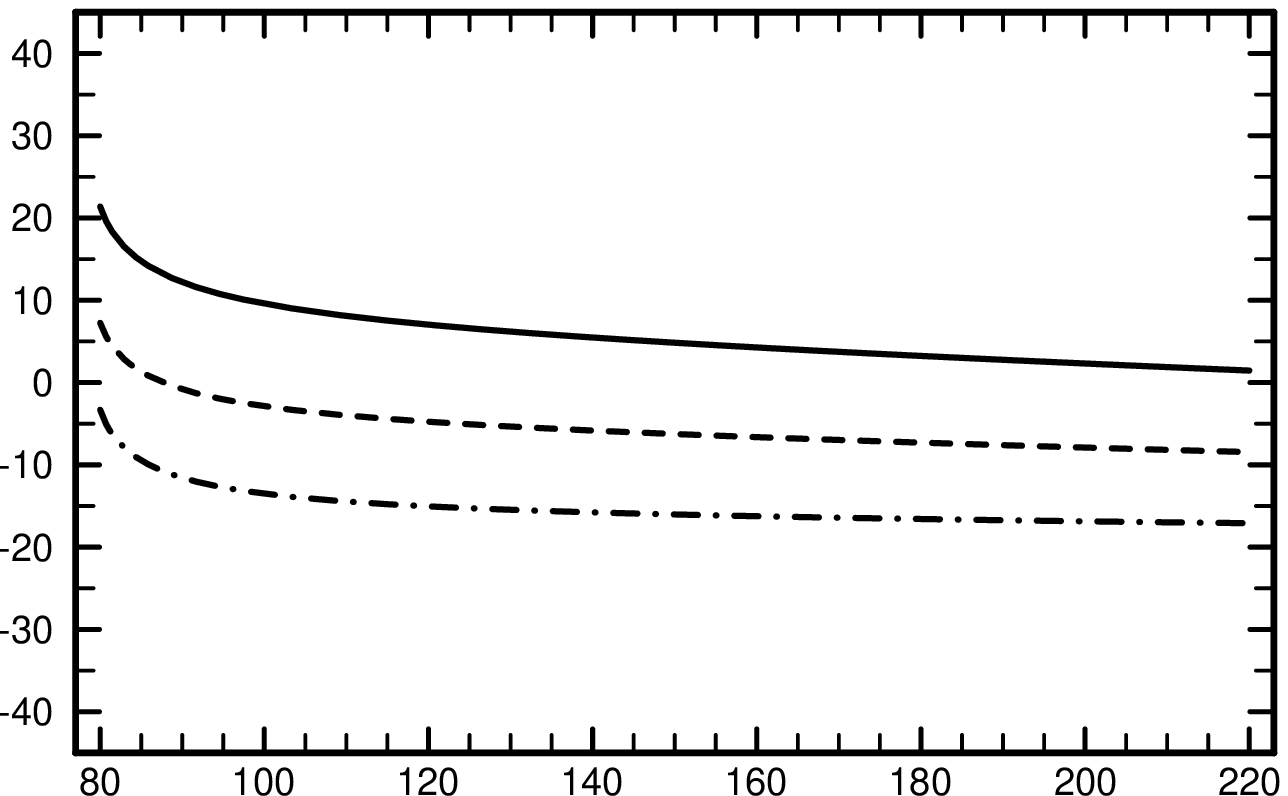,height=6.6cm}}}
\put(114,2){\makebox(0,0)[tr]{{$\mst{1}$~[GeV]}}}
\put(7,72){\makebox(0,0)[bl]{{\%}}}
\end{picture}}
\refstepcounter{figure}     
\bce{\large{\bf Fig.~\arabic{figure}}}\ece
\end{minipage}


\vspace{1cm}
\noi
\begin{minipage}[b]{125mm} {\setlength{\unitlength}{1mm} 
\begin{picture}(130,85)
\put(6,4){\mbox{\epsfig{figure=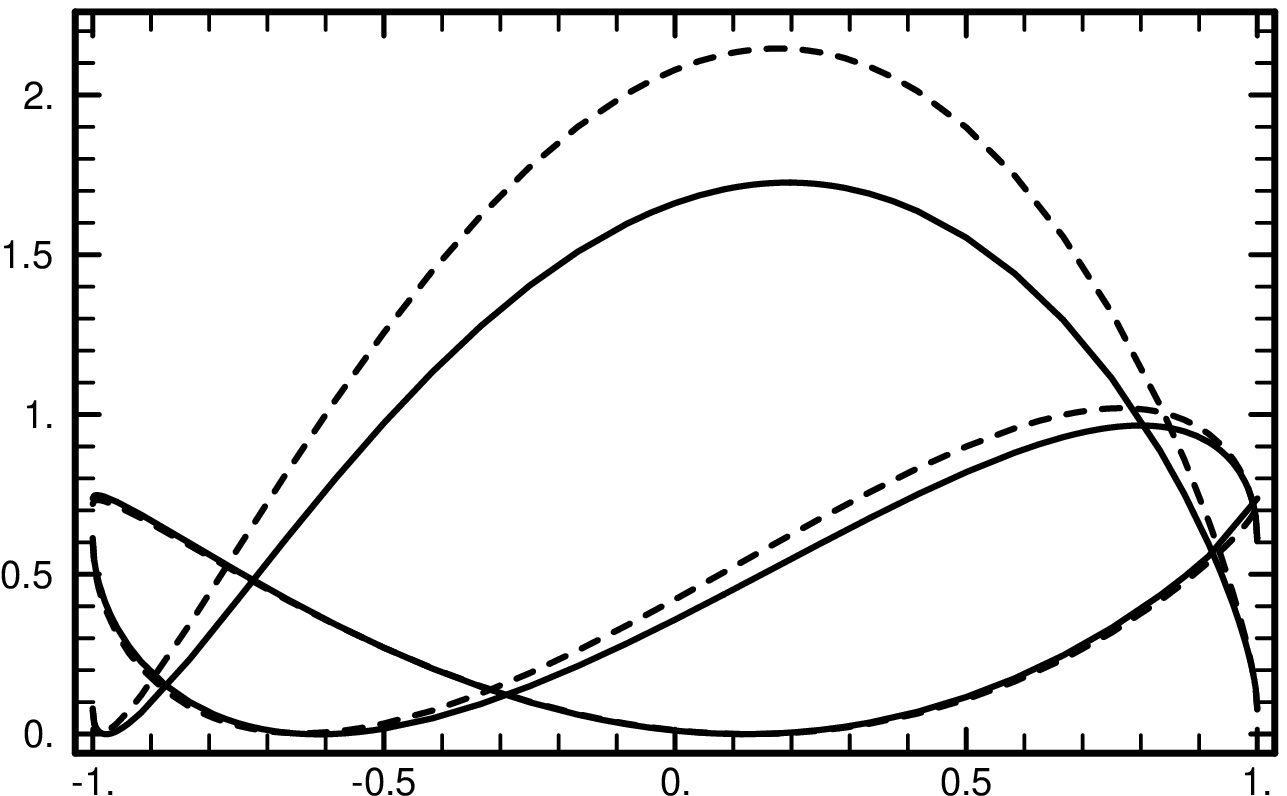,height=6.6cm}}}
\put(114,2){\makebox(0,0)[tr]{{$\cth_{\st}$}}}
\put(8,72){\makebox(0,0)[bl]{{$\Gamma$~[GeV]}}}
\put(86.5,13.5){
  \makebox(0,0)[br]{{\small  (a)}}}
\put(75,57){
  \makebox(0,0)[bl]{{\small  (c)}}}
\put(65,23.5){
  \makebox(0,0)[br]{{\small  (b)}}}
\end{picture}}
\refstepcounter{figure}     
\bce{\large{\bf Fig.~\arabic{figure}}}\ece
\end{minipage}


\vspace{1cm}
\noi
\begin{minipage}[b]{125mm} {\setlength{\unitlength}{1mm} 
\begin{picture}(130,85)
\put(6,4){\mbox{\epsfig{figure=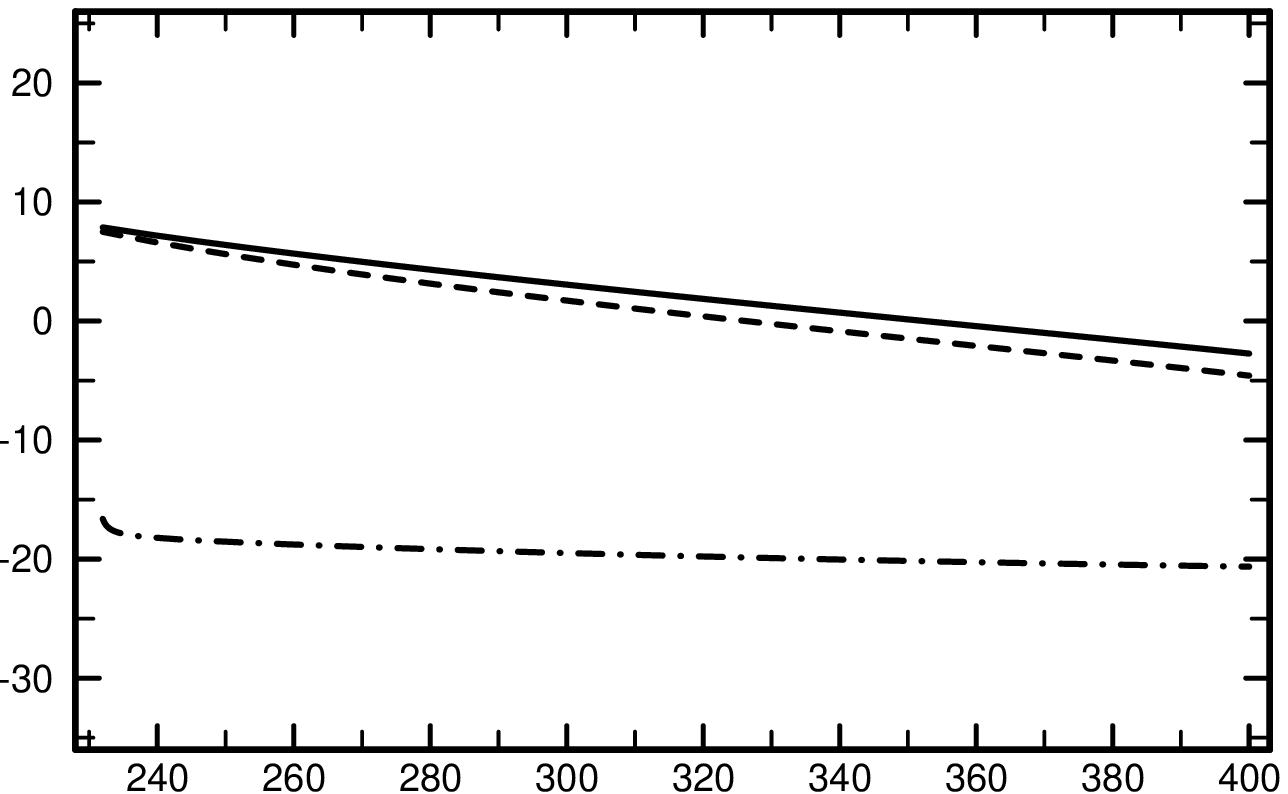,height=6.6cm}}}
\put(114,2){\makebox(0,0)[tr]{{$\mst{1}$~[GeV]}}}
\put(7,72){\makebox(0,0)[bl]{{\%}}}
\end{picture}}
\refstepcounter{figure}     
\bce{\large{\bf Fig.~\arabic{figure}}}\ece
\end{minipage}


\vspace{1cm}
\noi
\begin{minipage}[b]{125mm} {\setlength{\unitlength}{1mm} 
\begin{picture}(130,85)
\put(6,4){\mbox{\epsfig{figure=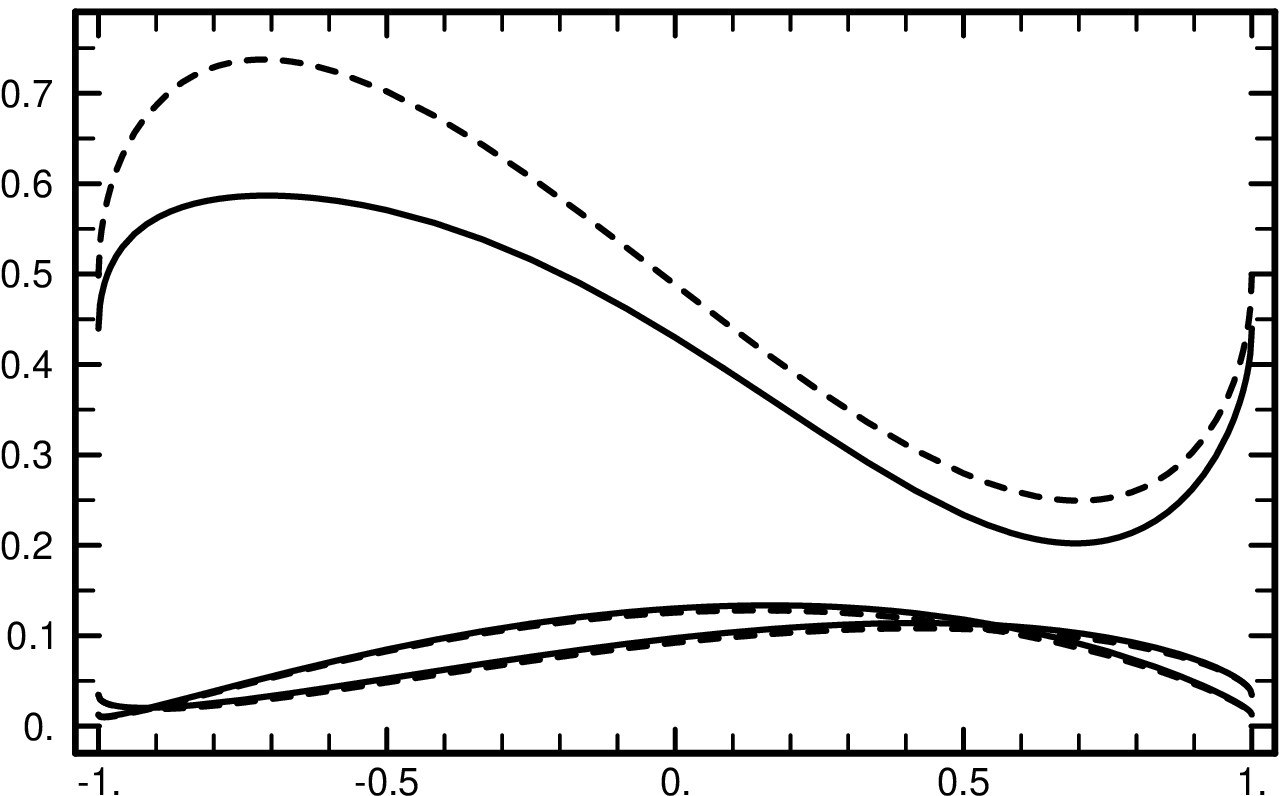,height=6.6cm}}}
\put(114,2){\makebox(0,0)[tr]{{$\cth_{\st}$}}}
\put(7,72){\makebox(0,0)[bl]{{$\Gamma$~[GeV]}}}
\put(52,20){
  \makebox(0,0)[br]{{\small (a)}}}
\put(68,44){
  \makebox(0,0)[bl]{{\small (c)}}}
\put(64,16){
  \makebox(0,0)[tl]{{\small (b)}}}
\end{picture}}
\refstepcounter{figure}     
\bce{\large{\bf Fig.~\arabic{figure}}}\ece
\end{minipage}

\end{center}
\end{document}